\documentstyle[12pt,epsfig]{article}

\def\Z{{\bf Z}}

\def\J{{\cal J}}

\date{}

\begin{document}

\title{{\LARGE\sf Are There Incongruent Ground States in $2D$ 
Edwards-Anderson Spin Glasses?}}
\author{
{\bf C. M. Newman}\thanks{Research partially supported by the 
National Science Foundation under grants 
DMS-98-02310 and DMS-01-02587.}\\
{\small \tt newman\,@\,courant.nyu.edu}\\
{\small \sl Courant Institute of Mathematical Sciences}\\
{\small \sl New York University}\\
{\small \sl New York, NY 10012, USA}\\
\and
{\bf D. L. Stein}\thanks{Research partially supported by the 
National Science Foundation under grants DMS-98-02153 and DMS-01-02541.}\\
{\small \tt dls\,@\,physics.arizona.edu}\\
{\small \sl Depts.\ of Physics and Mathematics}\\
{\small \sl University of Arizona}\\
{\small \sl Tucson, AZ 85721, USA}
}

\maketitle

\begin{abstract}
We present a detailed proof of a previously announced result 
\cite{NS2D00} supporting the absence of multiple (incongruent) ground state
pairs for $2D$ Edwards-Anderson spin glasses (with zero external field and,
e.g., Gaussian couplings): if two ground state pairs (chosen from
metastates with, e.g., periodic boundary conditions) on $\Z^2$ are
distinct, then the dual bonds where they differ form a {\it single\/}
doubly-infinite, positive-density domain wall. It is an open problem to
prove that such a situation cannot occur (or else to show --- much less
likely in our opinion --- that it indeed does happen) in these models. Our
proof involves an analysis of how (infinite-volume) ground states change as
(finitely many) couplings vary, which leads us to a notion of
zero-temperature {\it excitation metastates\/}, that may be of independent
interest.
\end{abstract}


\small
\renewcommand{\baselinestretch}{1.25}
\normalsize

\section{Introduction}
\label{sec:intro}

The decades-old challenge of understanding the physical nature of
laboratory spin glasses and the mathematical nature of spin glass models at
low temperature continues. It is a paradigm of the wider effort to analyze
the many novel features that occur in disordered systems generally. One can
only hope that this effort will achieve some fraction of the successes that
have been reached in understanding homogeneous systems --- in and out of
equilibrium --- and that are epitomized by the work of Joel Lebowitz and
his many collaborators.  It is indeed an honor to contribute to this
celebration of Joel's first $70$ years; may he live to $120$.

Our focus here is entirely on the Edwards-Anderson (EA)~\cite{EA} model on
$\Z^d$, simplest of the short-ranged Ising spin glasses, with Hamiltonian
\begin{equation}
\label{eq:EA}
{\cal H}_{\cal J}(\sigma)= -\sum_{\langle x,y\rangle} J_{xy} \sigma_x 
\sigma_y\quad .
\end{equation}
Here ${\cal J}$ denotes a specific realization of the couplings $J_{xy} =
J_{\langle x,y\rangle}$, the spins $\sigma_x=\pm 1$ and the sum is over
nearest-neighbor pairs $\langle x,y\rangle$ only, with the sites $x,y$ on
the square lattice $\Z^d$.  The $J_{xy}$'s are independently chosen from a
symmetric, continuous distribution with unbounded support, such as Gaussian
with mean zero; we denote by $\nu$ the overall disorder distribution for
${\cal J}$.

In this paper, we restrict attention entirely to ground states, and
further, to the lowest interesting dimension, $d=2$. Of course, for
$d=1$, and assuming as we do that the $J_{xy}$'s are continuously
distributed, it is easy to see that the multiplicity of
infinite-volume ground states is exactly two --- i.e., a single ground
state pair (GSP) of spin configurations related to each other by a
global spin flip --- since, in the absence of frustration, every bond
can be satisfied in a ground state.

We are interested in the question of
whether there are infinitely many {\it observable\/} GSP's.  By
``observable'' we mean that these states can be generated without using
special ${\cal J}$-dependent boundary conditions. This means that by using,
say, periodic boundary conditions on the $L \times L$ squares $S_L$ 
centered at the origin, for a sequence of $L$'s tending to infinity,
also chosen in a ${\cal J}$-independent way, the
corresponding sequence of finite-volume GSP's for the finite-volume
Hamiltonians ${\cal H}_{\cal J}^{(L)}$ (when restricted to a fixed, but
arbitrarily large window about the origin) will generate an empirical
distribution, i.e., a histogram, that in the limit is dispersed over many
GSP's.

\section{Main Result}
\label{sec:results}

\subsection{Preliminaries:  Metastates}
\label{subsec:meta}

To state a precise theorem about the GSP's that arise in this way, we need
to explain the notion of a metastate~\cite{NScp2,NS98,NSmf2,NSBerlin} in
this zero-temperature context.  We will do this in the briefest possible
way here, using empirical distributions, while delaying to later sections
of the paper a discussion of the fact that there are alternative
definitions giving rise to the same mathematical object. 

First, we note that for a given ${\cal J}$, with all couplings nonzero, a
GSP $\alpha$ may be identified with the collection of unsatisfied bonds,
which we regard as edges in the dual lattice.  Now suppose that $L_j \to
\infty$ is a sequence of scale sizes, not depending on ${\cal J}$, such
that for $\nu$-almost every ${\cal J}$, there is a probability measure
(called a metastate) $\kappa_{\cal J}$, defined on the configurations
$\alpha$ of GSP's on all of $\Z^2$, which is the limit of the empirical
distributions of the finite volume GSP's $\alpha_{\cal J}^{(L)}$ along the
sequence $L_j$ as follows: Let $D_1$ and $D_2$ be disjoint {\it finite\/}
sets of dual edges, let $A(D_1,D_2)$ denote the event that every edge in
$D_1$ is unsatisfied and every edge in $D_2$ is satisfied; let $F_{\cal
J}^{(M)} (D_1,D_2)$ denote the {\it fraction\/} of the indices $j \in
\{1,\dots,M\}$ such that all the edges of $D_1$ and $D_2$ are within the
square $S_{L_j}$ and such that the GSP $\alpha_{\cal J}^{(L_j)}$ obeys all
the requirements of $A(D_1,D_2)$; then for every such $D_1$ and $D_2$,
\begin{equation}
\label{eq:empiric}
\lim_{M\to\infty} F_{\cal J}^{(M)} (D_1,D_2)\, = \, 
\kappa_{\cal J} (A(D_1,D_2))\quad .
\end{equation}

Thus a metastate for $T=0$ is an ensemble of infinite-volume GSP's that
describes the asymptotic fractions of squares, along a subsequence $L_j$,
for which the various GSP's are observed (when restricted to windows of
fixed, but arbitrarily large, size) within the finite-volume systems.  It
can be shown by compactness arguments~\cite{NSmf2,NSBerlin} that such
subsequences $L_j$ exist; in fact every subsequence has such an $L_j$ as a
further sub-subsequence. Although it is a reasonable conjecture that any
two metastates are in fact the same for almost every ${\cal J}$, no general
result has been proved. However, this would be an immediate corollary of
the following conjecture, at least for $d=2$, which would also imply that
{\it the\/} metastate is supported on a {\it single\/} GSP for almost every
${\cal J}$.  We note that recent numerical results are consistent with the
existence of only a single GSP in two dimensions \cite{M99,PY99}.

\medskip

{\bf Conjecture 1.\/} Let ${\cal J}$ be chosen from the disorder
distribution $\nu$ and let $\alpha$ and $\beta$ be GSP's chosen
independently from $d=2$ periodic boundary condition metastates, $\kappa_{\cal J}$ and
$\kappa'_{\cal J}$ (coming from subsequences $L_j$ and $L'_k$). Then, with
probability one, $\alpha = \beta$.

\subsection{Theorem}
\label{subsec:theorem}

The main result of this paper is the proof of the following theorem,
which we regard as partial verification of the above Conjecture ---
see the Remark below. Equality of two GSP's, $\alpha$ and $\beta$, is
of course equivalent to the vanishing of the symmetric difference
$\alpha \Delta \beta$, the collection of bonds that are satisfied in
one of the two GSP's and unsatisfied in the other.  It is not hard to
show (see Proposition 1 below) that, at least for periodic boundary
conditions, the symmetric difference must consist either of a single
domain wall (i.e., a doubly-infinite self-avoiding path in the dual
lattice) with strictly positive density or else multiple
nonintersecting domain walls which have altogether strictly positive
density, but may have zero density individually.  A priori, we felt
(and still feel) that on a heuristic level, the former scenario for
GSP multiplicity is the less plausible of the two.  The next theorem
rigorously eliminates the latter scenario.

\medskip

{\bf Theorem 1.\/} Let ${\cal J}$ be chosen from the disorder distribution
$\nu$ and let $\alpha$ and $\beta$ be GSP's chosen independently from $d=2$
periodic boundary condition metastates, $\kappa_{\cal J}$ and $\kappa'_{\cal J}$ (coming
from subsequences $L_j$ and $L'_k$). Then, with probability one, either
$\alpha = \beta$ or else $\alpha \Delta \beta$ is a single domain wall with
strictly positive density.

\medskip

{\bf Proof.\/} This theorem will be an immediate consequence of three
propositions, given in Section~4 of the paper.

\medskip

{\bf Remark.\/} Although Theorem~1 does not eliminate the scenario of
multiple GSP's whose symmetric differences are {\it single\/} positive
density domain walls, we suspect that such domain walls do not in fact
occur. The proof of Theorem~1 is based on showing that the presence of
two or more $\alpha \beta$ domain walls would create an instability
for both $\alpha$ and $\beta$ with respect to the flip of a large
droplet whose boundary consists of two long segments from adjacent
domain walls, connected by two short ``rungs'' between the walls. The
stability of $\alpha$ and $\beta$ to such flips is controlled by the
infimum $E'$ of the necessarily positive rung energies --- see
Equation~(\ref{eq:rungenergy}). Proposition~3 of Sect.~4 proves
instability by showing that $E'=0$, while Proposition~2 there shows
that such unstable GSP's cannot actually occur with nonzero
probability. If there were a single domain wall, it would be natural
to expect that, like the rungs in Proposition~3, the ``pseudo-rungs''
that connect sections of the domain wall that are close in Euclidean
distance, but greatly separated in distance along the domain wall,
could also have arbitrarily low positive energies.  If these
pseudo-rungs connected long pieces of the domain wall containing some
fixed bond (and we emphasize that these properties have not been
proved), then single domain walls would be ruled out by an analogue of
Proposition~2.  The consequence would be that the periodic boundary
condition~metastate in the $2D$ EA Ising spin glass would be unique
and supported on a single GSP.

\subsection{Extension to Other Boundary Conditions}
\label{subsec:otherbcs}

The restriction to periodic boundary conditions in Theorem~1 can in fact be relaxed to
allow other boundary conditions that {\it do not depend on $\J$\/}. For boundary conditions such as
antiperiodic that are flip-related to periodic ones, nothing needs to be
done, since they yield the same metastate --- see Section IV of \cite{NS98}.

To explain how other boundary conditions can be handled, we begin by noting that the
significance of periodic boundary conditions is that they yield translation-invariance
of various infinite-volume objects, which in turn is a crucial ingredient
in the propositions of the next section.  With periodic boundary conditions,
translation-invariance is already valid for finite volume. For example,
from the random pair $({\J},\alpha_{\J}^{(L)})$, the finite dimensional
distributions of finitely many coupling values and finitely many bond
satisfaction variables are unchanged under translation by $y$, as long as
$y$ does not translate any of the finitely many bonds in question beyond
$S_L$. On the other hand, in the spirit of the empirical distribution
construction of the metastate described above, one could rather consider
the random pair $({\J},\alpha_{\J}^{({\cal L})})$, with ${\cal L}$ chosen,
uniformly at random, from ${L_1,\dots,L_M}$.  In that case, there is in a
certain sense only {\it approximate\/} translation invariance for finite
$M$, since the bonds typically do get translated out of $S_{L_j}$ for small
$j$. But full translation-invariance is restored in the limit
$M\to\infty$. 

For non-periodic, but still $\J$-independent,
boundary conditions, one can somewhat
similarly obtain infinite-volume translation-invariance, as follows.  For
each $L$ and $x$, let $\alpha_{\J}^{(L,x)}$ denote the GSP in the
translated square $S_L+x$ with some $\J$-independent boundary condition,
such as free or
plus. Next, let ${\cal X}(L)$ denote a uniformly random site in
$S_{L'(L)}$, where the deterministic $L'(L)\to \infty$ with, say,
$L-L'(L)\to \infty$ (e.g., $L'(L) = \sqrt{L}$).  Then the random pair
$({\J},\alpha_{\J}^{(L,{\cal X}(L))})$ or, alternatively,
$({\J},\alpha_{\J}^{({\cal L},{\cal X}({\cal L}))})$, has approximate
translation-invariance, which becomes exact as $L\to\infty$, or, alternatively,
$M\to\infty$. Using such an ``average over translates'' construction, one
can obtain metastates coming from, e.g., free or plus boundary conditions, for which the
analogue of Theorem~1 will be valid.  Such averaging over translates can
also be used to obtain translation-invariance for the extended notions of
metastates we describe next.

\section{The Excitation Metastate}
\label{sec:excitation}

An important part of the proof of Theorem~1 is based on extending the
notion of metastates so as to describe how a given GSP changes as the
couplings in ${\cal J}$ vary. Of course, if Conjecture 1 were true, then,
at least for $d=2$, there would be, for almost every ${\cal J}$, a GSP
$\alpha_{\cal J}$, uniquely determined as being the one on which the
periodic boundary condition metastate is supported; thus one would know how
$\alpha_{\cal J}$ changes even when {\it infinitely\/} many of the
couplings in ${\cal J}$ vary. But in general, since there might be many
GSP's and perhaps even many metastates, it is not so obvious how to
formulate the dependence of a given GSP in the support of a metastate even
on {\it finitely\/} many couplings.

Neither the {\it statement\/} of Theorem 1 nor that of our three main
propositions requires this extension of metastates, but it will be needed
for the proofs of the latter two of the main propositions. This extension
will be presented in detail in Section~5 of the paper, but we present a
short exposition here, since it seems to be of independent
interest. Roughly speaking, the extension requires that we keep track of
not only the GSP itself, but also of all its excitations in which finitely
many spins are forced to take specified values, modulo a global flip.  We
note that recent numerical studies of spin glasses have analyzed
excitations induced in this way \cite{KM} and in more novel ways
\cite{PY00}.  There are two types of information about our excitations that
one might wish to keep track of: (a) the minimum energy cost required to
force the spins, and (b) the pair of spin configurations that does the
minimizing --- i.e., the excited state. It actually suffices to keep track
only of (a), but it is perhaps conceptually simpler to keep track of (b) as
well, and we will take that tack.

Suppose $A$ is a finite subset of $\Z^2$ (in this discussion, we only take
$d=2$ for convenience), $\eta$ is a spin configuration on $A$ and $L$ is
sufficiently large so that $A \subset S_L$.  We denote by
$\alpha_{\J}^{A,\eta,(L)}$ the pair of periodic boundary condition spin
configurations on $S_L$ with minimum energy {\it subject to the constraint
that they equal $\pm \eta$ on $A$\/}. If $A$ is empty or a singleton site,
this is just the ordinary finite-volume ground state $\alpha_{\J}^{(L)}$.
We also define the excitation energy $\Delta E_{\J}^{A,\eta,(L)}$ to be the
energy of $\alpha_{\J}^{A,\eta,(L)}$ minus the ground state energy of
$\alpha_{\J}^{(L)}$.  Let $B$ be a finite set of bonds $b = \langle
x,y\rangle$ and let $\J^B$ denote a realization of the couplings $J_b$ for
all $b \in B$. To see how $\alpha_{\J}^{(L)}$ and eventually $\alpha_{\J}$
varies with $\J^B$ when all other couplings are fixed, we begin by letting
$A = A(B)$ denote the set of sites that are endpoints of bonds in $B$ and
considering the excitation energies $\Delta E_{\J}^{A,\eta,(L)}$ and
corresponding excited states $\alpha_{\J}^{A,\eta,(L)}$, for all possible
spin configurations $\eta$ on $A$. We also define
\begin{equation}
\label{eq:localH}
{\cal H}_{\J^B}(\eta)= -\sum_{{\langle x,y\rangle} \in B} 
J_{xy}^B \eta_x\eta_y \,\, ,\,\,\,\,
{\cal H}_{\J}(\eta;B)= -\sum_{{\langle x,y\rangle} \in B}
J_{xy} \eta_x\eta_y \,\, ,
\end{equation}
and denote by $\J[\J^B]$ the coupling configuration in which
each coupling $J_b$ of $\J$ with $b \in B$ is replaced by $J_b^B$
and all other couplings are left unchanged. Then,
for fixed $\eta$, 
$\alpha_{\J[\J^B]}^{A,\eta,(L)}$ does not depend on $\J^B$ and
\begin{eqnarray}
\label{eq:engydiff}
{\cal H}_{\J[\J^B]}^{(L)}(\alpha_{\J[\J^B]}^{A,\eta,(L)})-
{\cal H}_{\J[\J^B]}^{(L)}(\alpha_{\J[\J^B]}^{A,\eta',(L)})
\, = \, 
{\cal H}_{\J[\J^B]}^{(L)}(\alpha_{\J}^{A,\eta,(L)})
&-&
{\cal H}_{\J[\J^B]}^{(L)}(\alpha_{\J}^{A,\eta',(L)})
\nonumber
\\
= \, ({\cal H}_{\J^B}(\eta)-{\cal H}_{\J}(\eta;B))-
({\cal H}_{\J^B}(\eta')-{\cal H}_{\J}(\eta';B)))
&+&\Delta E_{\J}^{A,\eta,(L)}-\Delta E_{\J}^{A,\eta',(L)}\, .
\end{eqnarray}

Note that $\Delta E_{\J}^{A,\eta,(L)}$ depends on $\J$ but not on $\J^B$
while ${\cal H}_{\J^B}(\eta)$ depends on $\J^B$ but not on $\J$.  Consider
now the finitely many functions, as $\eta$ varies on $A$,
\begin{equation}
\label{eq:partengy}
h_\eta^{(L)}({\J^B}) \equiv \Delta E_{\J}^{A,\eta,(L)}
+{\cal H}_{\J^B}(\eta)-{\cal H}_{\J}(\eta;B).
\end{equation}
These are affine functions of ${\J^B}$, and if we define
$\eta_{\J}^{*(L)}({\J^B})$ to be the $\eta$ that minimizes
$h_\eta^{(L)}({\J^B})$, it follows that
\begin{equation}
\label{eq:gdstdep}
\alpha_{\J[\J^B]}^{(L)} = \alpha_{\J}^{A,\eta_{\J}^{*(L)}({\J^B}),(L)}\, .
\end{equation}
When letting $L \to \infty$, we will do so for the ground state
$\alpha_{\J}$ and simultaneously for the excitation energies $\Delta
E_{\J}^{A,\eta}$ and excited states $ \alpha_{\J}^{A,\eta}$ for all choices
of finite $A$ and spin configurations $\eta$ on $A$; a superscript
$^\sharp$ will denote that collection of choices.  Of course, this needs
to be done via a metastate construction that extends the ``ground
metastate'' $\kappa_\J$ described earlier, to what we will call the {\it
excitation metastate} $\kappa_\J^\sharp$.  The excitation metastate is a
probability measure on infinite-volume excitation energies and states for
the given $\J$, $(\Delta E^\sharp, \alpha^\sharp)$, which includes the
ground metastate since the ground state $\alpha$ can be obtained by
restricting $\alpha^\sharp$ to $A$ being the empty set (or a singleton,
since we are dealing with periodic boundary conditions that do not break spin-flip
symmetry). To see how the ground state $\alpha$ changes to
$\alpha_{[\J^B]}$ when the couplings in a fixed {\it finite\/} $B$ vary, we
can then use the infinite-volume extensions of our last two displayed
equations (where ${\cal H}_{\J^B}(\eta)$ and ${\cal H}_{\J}(\eta;B)$ are as
before): 
\begin{equation}
\label{eq:partengyiv}
h_\eta({\J^B}) \equiv \Delta E^{A(B),\eta}
+{\cal H}_{\J^B}(\eta)-{\cal H}_{\J}(\eta;B),
\end{equation}
and
\begin{equation}
\label{eq:gdstdepiv}
\alpha_{[\J^B]} = \alpha^{A(B),\eta^{*}({\J^B})}\, ,
\end{equation}
where $\eta^{*}({\J^B})$ is the $\eta$ on $A(B)$ that minimizes 
$h_\eta({\J^B})$.
\medskip

\section{The Main Propositions}

\label{sec:threeprops}

In this section, we present the three central propositions leading
immediately to Theorem~1.  The proof of the first of these, a direct
application to spin glasses of general $2D$ percolation results of Burton
and Keane \cite{BK91}, will be given in this section.  The proof of the
second and third propositions will be given in Sect.~6.  We begin with a
somewhat more detailed discussion of ground metastates than given in the
last section.  For simplicity, we continue to restrict the discussion to
periodic boundary condition metastates, as in Sect.~2.

An (infinite-volume) ground state pair or GSP for a given coupling
realization $\J$ is a pair of spin configurations $\pm \sigma$ on $\Z^d$,
whose energy, governed by Eq.~(\ref{eq:EA}), cannot be lowered by flipping
any {\it finite\/} subset of spins.  That is, it must satisfy the
constraint
\begin{equation}
\label{eq:loop}
\sum_{\langle x,y\rangle\in {\cal C}}J_{xy}\sigma_x \sigma_y \ge 0
\end{equation}
along any closed loop ${\cal C}$ in the dual lattice.  Infinite-volume
ground states are always the limits of finite volume ground states, but, in
general, the finite-volume boundary conditions may need to be carefully
chosen, depending on $\J$ and/or the limiting ground state.  In a
disordered sytem, if there are many distinct GSP's for typical fixed ${\cal
J}$, then in general, as noted in \cite{NScp1}, the limit
$\lim_{L\to\infty}\alpha_{\J}^{(L)}$ doesn't exist, if the $L$'s are chosen
in a coupling-independent way.  This phenomenon was called {\it chaotic
size dependence\/} \cite{NScp1}.  The ground metastate, a
probability measure $\kappa_{\cal J}$ on the infinite-volume ground states
$\alpha_{\cal J}$, was proposed in \cite{NSmf2} as a means of analyzing the
way in which $\alpha_{\J}^{(L)}$ samples from its various possible limits
as $L\to\infty$.  (The metastate was introduced and defined for both zero
and positive temperatures, but we confine the discussion here to zero
temperature.)  The same metastate can be constructed by at least two
distinct approaches.  The first, introduced earlier by Aizenman and Wehr
(AW) \cite{AW90}, directly employs the randomness of the ${\cal J}$'s,
while the ``empirical distribution'' approach of \cite{NSmf2} and
subsequent papers was motivated by, but doesn't require, the potential
presence of chaotic size dependence for fixed ${\cal J}$.

The empirical distribution point of view (and its natural extension to
excitation metastates) will be the primary one used throughout this paper.
However, we briefly describe the AW construction, since it is the one that
directly gives, for, e.g., periodic boundary conditions, the translation
invariance that will be crucial in our first proposition; for more details
see \cite{AW90}.  Here one considers, for each $L$, the random pair
$({\J},\alpha_{\J}^{(L)})$ (where $\alpha_{\J}^{(L)}$ is the finite-volume
periodic boundary condition GSP obtained using the restriction ${\cal
J}^{(L)}$ of ${\cal J}$ to $S_L$), and takes the limit of the
finite-dimensional distributions along a ${\cal J}$-independent subsequence
of $L$'s, using compactness.  This yields a probability distribution ${\cal
K}$ on infinite-volume $(\J,\alpha)$'s which is translation invariant,
under simultaneous lattice translations of $\J$ and $\alpha$, because of
the periodic boundary conditions, and is such that the conditional
distribution ${\tilde \kappa}_{\cal J}$ of $\alpha$ given $\J$ is supported
entirely on GSP's for that $\J$.  The conditional distribution ${\tilde
\kappa}_{\cal J}$ is the AW ground metastate.

It is easy to show that there is sequential compactness leading to
convergence for ${\cal J}$-independent subsequences of $L$'s, as described
above.  We have conjectured \cite{NSBerlin} that all subsequence limits are
the same; i.e., that existence of a limit does not require taking a
subsequence.  Proving this conjecture remains an open problem.

The empirical distribution approach of \cite{NScp2,NSmf2,NSBerlin}, as
described in Sect.~2, takes a fixed ${\cal J}$ and, roughly speaking,
replaces the ``${\cal J}$-randomness'' used in the AW construction of
${\tilde \kappa}_{\cal J}$ with ``$L$-randomness'' --- i.e., with chaotic size
dependence.  The empirical distributions along a subsequence
$(L_1,L_2,\dots)$ are the measures
\begin{equation}
\label{eq:emp}
\kappa_{\cal J}^M =(1/M)\sum_{k=1}^M\delta_{\alpha_{\J}^{(L_k)}}\, ,
\end{equation}
where $\delta_\alpha$ denotes the Dirac delta measure at
the state $\alpha$ and 
where for convenience we regard the finite-volume GSP $\alpha_{\J}^{(L)}$
as defined in infinite volume by, e.g., taking all bonds outside $S_L$ as
satisfied. We say that $\kappa_{\cal J}^M$ has a limit $\kappa_{\cal J}$ if
the probability of any event $A(D_1,D_2)$ (that every edge in $D_1$ is
unsatisfied and every edge in $D_2$ is satisfied, where $D_1$ and $D_2$ are
disjoint {\it finite\/} sets of dual edges) converges to the
$\kappa_{\cal J}$-probability of that event.

It was shown in \cite{NSBerlin} that there exists a ${\cal J}$-independent
subsubsequence where the limits ${\tilde \kappa}_{\cal J}$ and
$\kappa_{\cal J}$ are the same.  For more details and proofs, see
\cite{NScp2,NSmf2,NSBerlin}.  Also see \cite{NS98} for additional
properties of the metastate, particularly invariance with respect to
gauge-related boundary conditions.

Before we state Proposition~1, some additional definitions are needed.
Consider a periodic boundary condition~metastate $\kappa_{\cal J}$ (in some
fixed dimension, not necessarily two) and two GSP's $\alpha$ and $\beta$
chosen from $\kappa_{\cal J}$.  Then their {\it symmetric difference\/}
$\alpha\Delta\beta$, as introduced in Sect.~2, is the set of edges in the
dual lattice ${\bf Z^d}^*$ that are satisfied in $\alpha$ and not $\beta$
or vice-versa.  If ${\cal B}$ is the graph whose edge set is
$\alpha\Delta\beta$ and whose vertices are all sites in ${\bf Z^d}^*$
touching $\alpha\Delta\beta$, then a {\it domain wall\/}, defined relative
to the two GSP's, is a cluster (i.e., a maximal connected component) of
${\cal B}$. (In two dimensions, according to Proposition~1, domain walls
are generically doubly-infinite self-avoiding paths in the dual lattice.)
The symmetric difference $\alpha\Delta\beta$ is the union of all
$\alpha\beta$ domain walls and may consist of a single domain wall or of
multiple domain walls that are site-disjoint and hence also edge-disjoint.

Two distinct GSP's $\alpha$ and $\beta$ are said to be {\it incongruent\/}
if $\alpha\Delta\beta$ has a well-defined nonvanishing density within the
set of all edges in ${\bf Z^d}^*$; if the density is zero, they are
regionally congruent.  We do not consider here the case where the density
is not well-defined; we will see from Proposition~1 that in fact this
cannot happen in two dimensions.  In Proposition~1, we will also see that,
if there are multiple GSP's, the ``observable'' ones are incongruent.  Our
primary interest is therefore in the question of existence of these
``physical'' incongruent states, which should be observable by using
coupling-independent boundary conditions.  As mentioned in Sect.~2,
incongruent states may consist of a single positive-density wall, or else
of multiple domain walls, which individually may or may not have positive
density, but collectively have strictly positive density.

In all our propositions, ${\cal J}$ is chosen from the disorder
distribution $\nu$ and then $\alpha$ and $\beta$ are GSP's chosen
independently from periodic boundary condition metastates $\kappa_{\cal J}$
and $\kappa'_{\cal J}$ (which may be the same), as described above.

\medskip

{\bf Proposition 1.\/} \cite{NS2D00,BK91} Distinct $\alpha$ and $\beta$
in any dimension must, with probability
one, be incongruent.  In two dimensions, all domain walls comprising
$\alpha\Delta\beta$ have the following properties with probability one: 
(i) they are infinite
and contain no loops or dangling ends; (ii) they cannot branch
and thus are doubly-infinite self-avoiding paths; (iii) they
together partition ${\bf Z}^2$ into at most two topological
half-spaces and/or a finite or infinite number of
doubly-infinite topological strips (that also cannot
branch --- i.e., each strip has two boundary domain walls
and exactly one neighboring strip or half-space
on each side).  (iv) Moreover, each domain wall has a well-defined
density and there cannot simultaneously be positive-density and
zero-density walls.

\medskip

{\bf Proof of Proposition 1.\/} Let us denote by ${\cal D}_{\cal J}$ the
probability measure on configurations of $\alpha\Delta\beta$ corresponding
to choosing $\alpha$ and $\beta$ independently from $\kappa_{\cal J}$ and
$\kappa'_{\cal J}$, and denote by ${\cal D}$ the measure then obtained by
integrating out the couplings ${\cal J}$ with respect to the disorder
distribution $\nu$.  We claim that ${\cal D}$ is translation-invariant. To
see this, begin with the translation-invariant measures on joint
configurations of couplings and GSP's ${\cal K}\,(=\nu \kappa_{\cal J})$
and ${\cal K}'\,(=\nu \kappa'_{\cal J})$ and note that the natural coupling
$\nu \kappa_{\cal J} \kappa'_{\cal J}$, a measure on $({\cal
J},\alpha,\beta)$ configurations, retains translation-invariance.  ${\cal
D}$ is then translation-invariant since it is just the distribution of
$\alpha\Delta\beta$ with $(\alpha,\beta)$ distributed as the marginal of
this coupled measure. The translation-invariance of ${\cal D}$ in turn
implies by the ergodic theorem with respect to ${{\bf
Z}^2}^*$-translations that any ``geometrically defined event'', such as a
bond belonging to a domain wall, occurs either nowhere or else with
strictly positive density.  This proves the first claim.

To prove property (i), we note that a domain wall taken from
$\alpha\Delta\beta$ separates regions in which the spins of $\alpha$ and
$\beta$ agree from regions where they disagree.  A domain wall therefore
cannot end at a point in any finite region.  To rule out loops, note that
the sum $\sum_{\langle x,y\rangle}J_{xy}\sigma_x \sigma_y$ along any such
loop must have opposite signs in the two GSP's, violating
Eq.~(\ref{eq:loop}), unless the sum vanishes.  But this occurs with zero
probability because the couplings are chosen independently from a
continuous distribution.

Claims (ii), (iii), and (iv) are proven in \cite{BK91}, using
percolation-theoretic arguments first presented in \cite{BK89}; we sketch
the arguments.  To prove (ii), suppose that a domain wall branches at some
site $z$ in the dual lattice.  (We note, although it's not needed for the
proof, that the number of branches emanating from $z$ must be even, again
because domain walls separate regions of spin configuration agreement from
regions of disagreement.  Hence the minimal branching at $z$ is four.)
None of these branches may intersect somewhere else, by property (i).  By
the translation-invariance of ${\cal D}$, there must then be a positive
density of branch points, so that the domain wall would have a treelike
structure.  That implies the existence of an $\epsilon>0$ such that the
boundary of $S_L$ is intersected by a number of distinct branches that
grows as $\epsilon L^2$ as $L\to\infty$, which is impossible.

The proof of (iii) uses a similar argument to rule out branching of the
strips --- see Theorem 2 of \cite{BK91} for details.  Property (iv) is not
needed for subsequent arguments, but is included for completeness; it is
proven in Theorem 4 of \cite{BK91} and follows readily from the properties
just proven.  If zero-density and positive-density clusters coexist, then
for some $p>0$, there is positive ${\cal D}$-probability that the origin of
the dual lattice is contained in a zero-density domain wall with an
adjacent wall of density at least $p$.  Let ${\cal S}_p$ be the set of all
walls with density greater than or equal to $p$.  Then there can be no more
than $(1/p)$ walls in ${\cal S}_p$.  The maximum number of walls of density
zero that are adjacent to walls belonging to ${\cal S}_p$ (i.e., if every
${\cal S}_p$-wall is surrounded by two zero-density walls whose other
adjacent wall does not belong to ${\cal S}_p$) is therefore $2/p$.  But
then the union of such zero-density walls has density zero and so the
probability of the event that the origin is contained in a zero-density
wall adjacent to a wall in ${\cal S}_p$ is zero, leading to a
contradiction. This completes the proof of the proposition.

\medskip

So the picture we now have of the symmetric difference $\alpha\Delta\beta$
is a union of one or more doubly infinite domain walls.  These domain
walls do not branch or have any internal loops, and they divide the plane
into strips or (if there are positive-density domain walls) half-planes.
In all cases where there is more than a single domain wall,
translation-invariance of ${\cal D}$ implies that distinct domain walls
mostly remain within an $O(1)$ distance of one another.  E.g., there can
be no ``hourglass'', ``martini glass'', etc., domain wall configurations;
these can be ruled out by arguments similar to those used in the proof of
part (ii) of Proposition~1.

The essential idea behind the proof of Theorem~1 is contained in the next
two propositions.  Before we state these propositions, we need to introduce
the notion of a ``rung'' between adjacent domain walls.  A rung ${\cal R}$,
defined with respect to $\alpha\Delta\beta$, is a path of edges in ${{\bf
Z}^2}^*$ connecting two distinct domain walls, with only the first and last
sites in ${\cal R}$ on any domain wall.  So ${\cal R}$ can contain only
edges that are {\it not\/} in $\alpha\Delta\beta$, and the corresponding
couplings are therefore either both satisfied or both unsatisfied in
$\alpha$ and $\beta$.  The energy $E_{\cal R}$ of ${\cal R}$ is defined to
be
\begin{equation}
\label{eq:rungenergy}
E_{\cal R}=\sum_{\langle xy\rangle \in {\cal R}}J_{xy}\sigma_x\sigma_y\ ,
\end{equation}
with $\sigma_x\sigma_y$ taken from $\alpha$ or equivalently $\beta$.  It
must be that $E_{\cal R}>0$ with probability one for the following reasons,
which we sketch here and make precise later in the proof of Proposition 2.
Suppose that a rung could be found with negative energy (there is zero
probability of a zero-energy rung); by translation-invariance there would
need to be many such rungs between some fixed pair of adjacent domain
walls.  Consider the ``rectangle'' formed by two such negative-energy rungs
and the connecting segments of the two adjacent domain walls.  The sum of
$J_{xy}\sigma_x\sigma_y$ along the couplings in the domain wall segments
would be positive in one GSP (say, $\alpha$), and would therefore be
negative in the other (say, $\beta$).  Therefore, the loop formed by the
boundary of this rectangle would violate Eq.~(\ref{eq:loop}) in GSP
$\beta$.

It is then natural to ask the deeper question of whether rung energies
along any strip are strictly bounded away from zero, or whether their
infimum is exactly zero.  Propositions~2 and 3 address this question.

\medskip

{\bf Proposition 2.\/} The rung energies $E_{{\cal R}'}$ between two fixed
(adjacent) domain walls cannot be arbitrarily small; i.e., there is zero
probability that $E'=\inf_{{\cal R}'}E_{{\cal R}'}=0$.

\medskip

{\bf Proposition 3.\/} There is zero probability that $E'>0$.

\medskip

The contradiction between Propositions 2 and 3 leads directly to Theorem~1.
These propositions will be proved in Section~\ref{sec:proofof23}.

\section{Transition Values and Flexibilities}
\label{sec:transition}

In this section, we present two auxiliary propositions. They will be used
in the next section to prove Propositions 2 and~3.  These auxiliary
propositions involve two notions, transition value and flexibility, that
arise in the analysis of how a GSP changes when a {\it single\/} coupling,
$J_b$, varies. Since this is a restricted case of the dependence of
$\alpha_{[\J^B]}$ on a finite collection $\J^B$ of couplings, we begin the
section by providing a more detailed exposition of the excitation metastate
than that given in Sect.~3 above.

Along with an empirical distribution construction of the excitation
metastate $\kappa_\J^\sharp$ as a probability measure, defined for
$\nu$-almost every $\J$, on configurations $(\Delta E^\sharp,
\alpha^\sharp)$ of excitation energies and states for the given $\J$, there
is an alternative AW-type construction, as follows. For each $L$, consider
$(\J, \Delta E_\J^{\sharp,(L)}, \alpha_\J^{\sharp,(L)})$, where $\Delta
E_\J^{\sharp,(L)}$ and $\alpha_\J^{\sharp,(L)}$ denote the excitation
energies and states in $S_L$, with periodic boundary conditions, when the
spin configuration on $A \subset S_L$ is constrained to be $\pm \eta$ (for
all allowed $A$'s and $\eta$'s).  As in the AW ground metastate
construction, one has sequential compactness of the corresponding
probability measures, ${\cal K}^{\sharp,(L)}$, leading to convergence of
the finite dimensional distributions (involving finitely many couplings,
finitely many finite $A$'s and finitely many $\eta$'s) to those of a
limiting translation-invariant measure ${\cal K}^\sharp$ on infinite-volume
configurations $(\J, {\Delta E^\sharp}, {\alpha^\sharp})$ along
deterministic subsequences of $L$'s.

The marginal distribution
of $\J$ from this ${\cal K}^\sharp$
is of course just $\nu$ and the conditional distribution
of $(\Delta E^\sharp, \alpha^\sharp)$ given $\J$ is then an
excitation metastate ${\tilde \kappa}_\J^\sharp$, which,
like in the ground metastate case, can be shown 
for $\nu$-almost every $\J$ to equal
the $\kappa_\J^\sharp$ constructed via empirical distributions,
as the limit along a subsubsequence of
\begin{equation}
\label{eq:exciteemp}
(1/M)\sum_{k=1}^M
\delta_{(\Delta E_\J^{\sharp,(L_k)}, \alpha_\J^{\sharp,(L_k)})}\, .
\end{equation}
The translation-invariance of ${\cal K}^\sharp$ follows, as usual, from the
periodic boundary conditions.  The relative compactness (tightness) for
${\alpha^{\sharp,(L)}}$ follows from the two-valuedness of spin
variables. Finally, the relative compactness (tightness) for ${\Delta
E^{\sharp,(L)}}$ follows from the trivial bound,
\begin{equation}
\label{eq:excitebound}
|\Delta E_{\J}^{A,\eta,(L)}| \le {\sum}_A '|J_{xy}|\, ,
\end{equation}
where $\sum_A '$ denotes the sum over bonds $\langle x,y\rangle$
with either $x$ or $y$ or both in $A$, together with the fact that
the distribution of the $J_{xy}$'s does not change with $L$.

As explained in Sect.~3, for a given $\J$, we can extract from 
$(\Delta E^\sharp, \alpha^\sharp)$ not only the GSP $\alpha$, but also
$\alpha_{[\J^B]}$, which describes how the GSP changes
when the couplings in a fixed finite set $B$ of bonds vary.
When $B$ consists of a single bond $b=\langle x,y\rangle$, 
we write $\alpha(K';b)$
for the ground state that results when $J_b$ is replaced by $K'$
with all other couplings of $\J$ left unchanged. It should be clear 
from Equations (\ref{eq:partengyiv}) and (\ref{eq:gdstdepiv})
that as $K'$ varies in $(-\infty,+\infty)$, the GSP $\alpha(K';b)$
changes exactly once (this is particularly easy to see in finite
volume and the property is preserved in the excitation metastate),
from its original configuration $\alpha$ when $K'= J_b$
to a new configuration 
\begin{equation}
\label{eq:gdstdepsb}
\alpha^b = \alpha^{\{x,y\},{\hat \eta}}\, ,
\end{equation}
where ${\hat \eta}$ is one of the two spin configurations on $\{x,y\}$
of opposite parity to the original GSP $\alpha$ (so that $\sigma_x \sigma_y$
is $+1$ in one of $\alpha$ and $\alpha^b$ and $-1$ in the other, or
equivalently $J_b$ is satisfied in one and unsatisfied in the other). 
We call the value of $K'$
where this change happens the transition value and denote it by $K_b$.

For a given $b$, the transition value $K_b$ and the {\it unordered\/}
set of two
GSP's $\{\alpha,\alpha^b\}$ do {\it not\/} depend on the value of $J_b$,
with all other couplings held fixed (again, this is clear for finite
volume, and is preserved in the limit). This means that with respect to
the probability measure ${\cal K}^\sharp$ on infinite-volume configurations
$(\J, {\Delta E^\sharp}, {\alpha^\sharp})$, {\it the random variables
$K_b$ and $J_b$ are independent\/}. 
The next proposition is an immediate consequence
of this independence.

\medskip

{\bf Proposition 4.\/} With probability one, no coupling $J_b$ is
exactly at its
transition value $K_b$.

\medskip

{\bf Proof of Proposition 4.\/} From the independence of $J_b$ and $K_b$, and the continuity
of the distribution of $J_b$, it follows that there is probability zero
that $J_b - K_b =0$.

\medskip

As in the proof of the last proposition, we continue to work on the
probability space of $(\J, {\Delta E^\sharp}, {\alpha^\sharp})$
configurations with probability measure ${\cal K}^\sharp$.  When the value
of $J_b$ is moved from its original value past the transition value $K_b$,
the change from the original ground state of $\alpha$ to the new ground
state, and originally excited state, of $\alpha^b$ may involve the flipping
of a finite droplet (region of $\Z^2$) or one or more infinite 
droplets. Thus the symmetric difference $\alpha \Delta \alpha^b$,
representing the dual bonds which change from satisfied to unsatisfied or
vice-versa, may consist of a single finite loop or else of one or more
infinite disconnected paths, but in all cases some part must pass through
$b$ since its satisfaction status clearly changes.  To help analyze what
other bonds $\alpha \Delta \alpha^b$ may or may not pass through, we
introduce the notion of flexibility.  The flexibility of a bond $b=\langle
x,y\rangle$ is defined as
\begin{equation}
\label{eq:flex}
F_b \equiv |K_b-J_b| = (1/2)\, |\Delta E^{\{x,y\},{\hat \eta}}|
\end{equation}
and thus is proportional to the excitation energy needed to flip
the relative sign of the spins at $x$ and $y$; it is a measure of
the stability of the ground state $\alpha$ with respect to fluctuations
of the single coupling $J_b$.

\medskip

{\bf Proposition 5.\/} For two bonds $a$ and $b$,
there is zero probability that $F_b > F_a$ and simultaneously
$\alpha \Delta \alpha^a$ passes through $b$.

\medskip

{\bf Proof of Proposition 5.\/} For finite $L$, and a bond $e$ in $S_L$,
let us denote by $F_e^{(L)}\equiv |J_e - K_e^{(L)}|$ the finite-volume
flexibility. Now $F_e^{(L)}$ is clearly the minimum, over all droplets in
$S_L$, with periodic boundary conditions,
whose boundary passes through $e$, of (half
the) droplet flip energy cost in the GSP $\alpha^{(L)}$. Since this is the
case for both $e=a$ and $e=b$, it is an immediate consequence that the
finite-volume droplet boundary $\alpha^{(L)} \Delta \alpha^{a,(L)}$ cannot
pass through $b$ if $F_b^{(L)} > F_a^{(L)}$. After $L \to \infty$, the
characterization of $F_e$ as a minimum over finite droplets may be lost,
but we claim that the conclusion of the proposition still holds. This is
because, although the convergence of ${\cal K}^{\sharp,(L)}$ along a
subsequence to ${\cal K}^\sharp$ is not sufficient to imply, e.g., that
the probability of $F_b^{(L)} > F_a^{(L)}$ converges along the
subsequence to the limiting probability of $F_b > F_a$, it is sufficent to
imply that the probability of the event in the proposition is less than or
equal to the the lim~inf of the (zero) probability of the corresponding
finite-volume events. This completes the proof of the proposition. 

\medskip

\section{Proof of Propositions 2 and 3} 
\label{sec:proofof23}

{\bf Proof of Proposition 2.\/} Suppose that there are two adjacent domain
walls from the GSP's $\alpha$ and $\beta$, $W_1$ and $W_2$, with $W_1$
passing through the origin of the dual lattice, and suppose further that
the infimum $E'$ of rung energies $E_{{\cal R}'}$ for rungs ${\cal R}'$
between $W_1$ and $W_2$ is zero. Our object is to prove that this event has
zero probability. If the probability is nonzero, then for every $\epsilon
>0$ there is some $\ell(\epsilon) < \infty$ so that, with nonzero
probability, there is a rung ${\cal R}'$ between $W_1$ and $W_2$, with the
property ${\cal P}(\epsilon)$, that its length, defined as the number of
bonds, is below $\ell(\epsilon)$ and its energy $E_{{\cal R}'}$ is below
$\epsilon$. But then, by translation-invariance and the lemma given right
after this proof, there must, with nonzero probability, be infinitely many
such rungs with property ${\cal P}(\epsilon)$ with starting points on $W_1$
in {\it both\/} directions from the origin along $W_1$.  Thus we can find
two such rungs ${\cal R}$ and ${\cal R}'$, one in each direction, and
sufficiently far apart that they do not touch each other.

Consider the ``rectangular'' region of $\Z^2$ whose
boundary is the union of these two rungs and the connecting segments,
$C_1$ and $C_2$ of
$W_1$ and $W_2$. The energy cost of flipping the spins in this region
in $\alpha$ (respectively, in $\beta$) is $+E(C_1,C_2)+E_{{\cal R}}
+E_{{\cal R}'}$ (respectively, $-E(C_1,C_2)+E_{{\cal R}}
+E_{{\cal R}'}$). Both these quantities must be positive since
both $\alpha$ and $\beta$ are GSP's; hence $|E(C_1,C_2)|$ is bounded
by $E_{{\cal R}}+E_{{\cal R}'} < 2 \epsilon$ and the energy costs
in both ground states are bounded by $4\epsilon$. This implies
that every bond $b$ that $W_1$ (or $W_2$) passes through has
flexibility less than $2 \epsilon$. Since $\epsilon$ is arbitrary,
the flexibilities must be zero, but that would contradict 
Proposition~4. This, together with the following lemma, completes the proof.

\medskip

{\bf Lemma 1.\/} Suppose ${\cal P}$ is a translation-invariant
property of rungs, e.g., the property that the rung energy is below
a certain value and/or the rung length is below a certain value.
There is zero probability that there exist two adjacent domain walls,
$W_1$ and $W_2$, such that the set of starting points on $W_1$ of rungs
between $W_1$ and $W_2$ that satisfy ${\cal P}$ is nonempty without
being doubly infinite, i.e., along {\it both\/} directions of $W_1$.

\medskip 

{\bf Proof of Lemma 1.\/} The proof is based entirely on the translation
invariance of the measure ${\cal K}^\sharp$. Suppose the claim of the lemma
is false. Then for each site $x$ in the dual lattice, there is nonzero
probability for the event $A_x$ that there is a domain wall $W$ passing
through $x$ and an adjacent wall $W'$ such that $x$ is the {\it last\/}
site in one of the two directions along $W$ such that there is a rung from
that site to $W'$ satisfying ${\cal P}$. Since every domain wall has two
directions and at most two adjacent domain walls, {\it there can be at most
four sites on any domain wall for which this event occurs\/}. Every domain
wall that intersects the square $S_L$, sitting inside the infinite lattice,
much touch the boundary of the square and thus there are at most $cL$ such
domain walls for some constant $c<\infty$, and consequently at most $4cL$
sites $x$ in $S_L$ for which $A_x$ occurs.  But by the ergodic theorem for
spatial translations, there is nonzero probability that the number of such
sites exceeds $c'L^2$ for some constant $c'>0$. This contradiction
completes the proof.

{\bf Proof of Proposition 3.\/} 

For the proof, we need the notion of a ``super-satisfied'' bond $b=\langle
x,y \rangle$.  It is easy to see, for a given $\J$, that $b$ is satisfied
in every ground state if $|J_{xy}|>$min$\{M_x,M_y\}$, where $M_x$ is the
sum of the three other coupling magnitudes $|J_{xz}|$ touching $x$, and
$M_y$ is defined similarly.  Such a bond or its dual, called
super-satisfied, clearly cannot be part of a domain wall between any two
GSP's.

As in the proof of Proposition~1, but using the excitation metastates
$\kappa_\J^\sharp$ and ${\kappa'}_\J^{\sharp}$ that extend the ground
metastates from which $\alpha$ and $\beta$ are chosen, we work in the
probability space with the coupled measure $\nu \kappa_\J^\sharp
{\kappa'}_\J^{\sharp}$. On this space, we can consider the modified ground
states $\alpha_{[\J^B]}$ and $\beta_{[\J^B]}$ as any finitely many
couplings are varied as well as the transition values and flexibilities
for both $\alpha$ and $\beta$ for all bonds $b$.

Now suppose that the rung energy infimum $E'$ between some pair $W_1,W_2$
of domain walls satisfies $E'>0$ with positive probability; we show this
leads to a contradiction.  First we find, as in Fig.~1, a rung ${\cal R}$
and two dual bonds $b_1,b_2$ whose locations on $W_1$ are respectively in
opposite directions from the starting site of ${\cal R}$, and such that
$E_{\cal R}-E'$, which we denote by $\delta$, is strictly less than the
flexibility values for both $\alpha$ and $\beta$ of both $b_1,b_2$.  The
existence with positive probability of such an ${\cal R}$, $b_1$ and $b_2$
follows from the non-vanishing of flexibilities given by Proposition~4 and
translation-invariance (e.g., Lemma~1).

\begin{figure}
\centerline{\epsfig{file=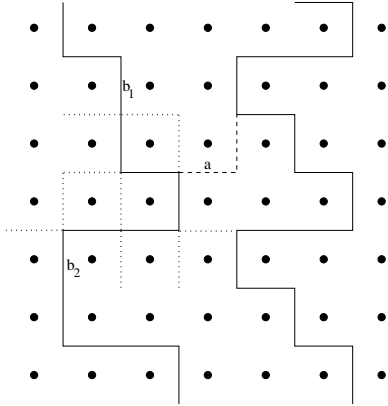,width=2.0in}}
\vspace{0.2in}
\caption{A rung ${\cal R}$ with $E_{\cal R} = E' +\delta$.  The dots are
sites in ${\bf Z}^2$, and bonds are drawn in the dual lattice.  Two
domain walls are solid lines and ${\cal R}$ is the dashed line.  The
bonds $b_1$ and $b_2$ have flexibility $>\delta$.  The ten dotted line
bonds are super-satisfied.}
\label{fig:bounce}
\end{figure}

But we also want a situation, as in Fig.~1, where all the dual lattice
non-domain-wall bonds that touch $W_1$ between $b_1$ and
$b_2$, other than the first bond $a$ in ${\cal R}$, are
super-satisfied, and remain so regardless of changes of $J_a$
(by a bounded amount).  We will call these bonds, 
numbering ten in Fig.~1, the ``special'' bonds. How do we
know that such a situation will occur with nonzero probability?  
If necessary, we can first adjust the signs and then increase the magnitudes
(in an appropriate order) of the couplings of the special bonds,
so that they first
become satisfied and then super-satisfied. This can be done in an
``allowed'' way because of our assumption that the distribution of
individual couplings has unbounded support. Also, this can be done so
that  $\alpha_{[J^B]}$ and $\beta_{[J^B]}$ remain unchanged
from $\alpha$ or $\beta$, and without changing
$E_{\cal R}$, without decreasing any other $E_{{\cal R}'}$ (and thus
without changing $E'$ or $E_{\cal R}-E'=\delta$) and without
decreasing the flexibilities of $b_1$ or $b_2$. Starting from a nonzero
probability event, such an allowed change of finitely many couplings in
${\cal J}$ yields an event which still has nonzero probability.

Next, suppose we move $J_a$ toward its transition value $K_a$ by an amount
slightly greater than $\delta$.  
The geometry --- see, e.g., Fig.~1 --- and 
Proposition~5 forbid the replacement
of either $\alpha$ or $\beta$ by $\alpha^a$ or $\beta^a$, because it is
impossible, under the conditions given, for $\alpha \Delta \alpha^a$ or
$\beta \Delta \beta^a$ to connect to the end of bond $a$ touching $W_1$.
But this change of $J_a$ reduces $E_{\cal R}$ below $E_{{\cal R}'}$ for any
${\cal R}'$ not containing $a$, yielding a nonzero probability event that
contradicts translation-invariance (i.e., Lemma~1).  This completes the
proof.

\medskip

\newpage

\renewcommand{\baselinestretch}{1.0}

\end{document}